\documentclass[fleqn,10pt]{wlscirep}

\title{Balancing Speed and Coverage by Sequential Seeding in Complex Networks}

\usepackage{float}
\usepackage{graphicx}
\usepackage{array}
\usepackage{kpfonts}
\usepackage{lipsum}                     
\usepackage{xargs}                      
\usepackage{xcolor}  
\usepackage{todonotes}
\author[1,2,*]{Jarosław Jankowski}
\author[1]{Piotr Bródka}
\author[1]{Przemysław Kazienko}
\author[1,3]{Boleslaw K. Szymanski}
\author[1]{Radosław Michalski}
\author[1]{Tomasz Kajdanowicz}

\affil[1]{Wrocław University of Science and Technology, Department of Computational Intelligence, Wrocław, 50-370, Poland}
\affil[2]{West Pomeranian University of Technology, Department of Computer Science and Information Technology, Szczecin, 71-210, Poland}
\affil[3]{Rensselaer Polytechnic Institute, Department of Computer Science, Troy, 12180, United States}

\affil[*]{jjankowski@wi.zut.edu.pl}


\begin{abstract}
Information spreading in complex networks is often modeled as diffusing information with certain probability from nodes that possess it to their neighbors that do not. Information cascades are triggered when the activation of a set of initial nodes – seeds – results in diffusion to large number of nodes. Here, several novel approaches for seed initiation that replace the commonly used activation of all seeds at once with a sequence of initiation stages are introduced. Sequential strategies at later stages avoid seeding highly ranked nodes that are already activated by diffusion active between stages. The gain arises when a saved seed is allocated to a node difficult to reach via diffusion. Sequential seeding and a single stage approach are compared using various seed ranking methods and diffusion parameters on real complex networks. The experimental results indicate that, regardless of the seed ranking method used, sequential seeding strategies deliver better coverage than single stage seeding in about 90\% of cases. Longer seeding sequences tend to activate more nodes but they also extend the duration of diffusion. Various variants of sequential seeding resolve the trade-off between the coverage and speed of diffusion differently.
\end{abstract}

\begin{document}

\flushbottom
\maketitle
%
%
\thispagestyle{empty}

\section*{Introduction}

The process of making the complex decisions is difficult, so it is often worth making partial decisions and to track their consequences before proceeding further. Such strategy was proven useful in areas such as: general theory of decision making \cite{Wald:1947,Siegmund:1985}, financial markets\cite{Cyert:1978,Prastacos:1983}, epidemiology\cite{PRICE01111968} and marketing \cite{Elberse:2003}. Here, we show that sequential, consecutive approach is also highly efficient in choosing the individuals, called seeds, that when activated will widely spread information or opinion in a social network. The current research on influence maximization and information spread in complex networks focuses mainly on single stage seed initiation. An exception is new product adaptation with early diffusion of product samples  \cite{Bass1969}$^{,}$\cite{Rogers2003} to benefit from consumer responses and product spread. The main challenge is finding a method for selection of seeds to maximize the final spread of information within the network. If the total number of seeds to be used is limited, e.g. due to restricted budget, a typical approach is to rank all nodes in the network according to some criteria, select top $n$ nodes as seeds and activate them at once to initiate the diffusion.

Influence maximization problem in complex networks was defined by Kempe\cite{Kempe:2003}. Analyses of various factors affecting the diffusion and social influence in complex networks include the efficiency of using different centrality measures for ranking influencers for selection\cite{Kiss2008}, impact of homophily for successful seeding,\cite{Nejad:2015} and heterogeneous thresholds on congestion, \cite{karampourniotis2015impact} finding the critical initiator fraction beyond which the cascade becomes global\cite{singh2013threshold} or importance of different network features in predicting spread.\cite{liu2012seeding} Selection of initial seeds was also analyzed, including incentives for innovators to start diffusion\cite{Lehmann:2006} and the multi-market entry perspective\cite{Libai:2005}. While most of research is related to marketing, the problem has been generalized to ranking nodes for target set selection in the domain of combinatorial optimization of theoretical computer science. \cite{Ackerman:2010}$^{,}$\cite{Ben-Zwi:2011}$^{,}$\cite{Chiang:2013} The influence maximization problem is also explored in physics from the perspective of diffusion in complex networks.\cite{Galstyan2009} Other studies discuss the role of communities\cite{He2015} and propose the influence maximization through optimal percolation\cite{Morone:2015}. Some initial research has been carried out to identify efficient seeds to initiate spread in temporal \cite{Michalski:2014}$^{,}$\cite{jankowski2013} and multilayer social networks. \cite{Michalski:2013}. Several comparative studies discuss seeding strategies\cite{Hinz:2011} and marketing word of mouth programs\cite{Libai:2013}. 

Typically, seeding strategies applied in viral marketing, social campaigns, political campaigns, diffusion of information, etc. are based on the assumption that selected seeds are initiated at once, and then the diffusion continues without any additional support \cite{hinz2011seeding}. Resent research takes into account adaptive approaches with two-stage stochastic model exploring the potential of neighboring nodes\cite{Seeman2013}, further extended towards more scalable approach\cite{Horel2015}. Other preliminary studies were performed to demonstrate the potential of using seeds after the first stage\cite{sela2015improving, zhang2015dynamic}. The potential of multi-period spraying algorithm for routing in delay-tolerant networks was also discussed\cite{Szymanski:2010}. Some authors noticed the benefit of not initiating a node that is likely to be activated by other seeds. Hence, they  proposed structural heuristics for static seed ranking that avoid choosing as a new seed (1) a node that is a neighbor of or close to the already chosen seed, \cite{Kitsak2010}$^{,}$\cite{Zhang2016} or (2) a node that is in a local cluster already containing another seed. \cite{Zhao2014}$^{,}$\cite{He2015}

Here, we introduce sequential seeding for influence maximization in complex networks and compare it to classical single stage approach. In our approach seeds are divided into several packages, and seeds from each package start either in fixed intervals or when diffusion stops for the previous package. We verify the performance of the proposed approach for different parameters related to network structure and characteristics of diffusion. Our results show that deploying the same number of seeds in stages as in at once approach yields larger spread, than achieved in a single stage seeding thanks to use of diffusion between stages. This improvement is however offset by the longer time needed to finish diffusion and to reach the maximal coverage. The best sequential seeding strategy depends on the relative importance of the spread coverage and the time available for spread execution.

A typical single seeding  approach starts with ranking all nodes according to some structural features ranging from a simple degree ranking \cite{hinz2011seeding} to more advanced solutions. \cite{Kitsak2010}$^{,}$\cite{Zhang2016}$^{,}$\cite{Zhao2014}$^{,}$\cite{He2015} Actual seed selection commences with initiating all $n$ top ranking nodes at once at the beginning of campaign. In contrast, sequential seeding utilizes the same ranking as the corresponding single stage method but initiates nodes that are top ranked and \textit{not yet initiated} in stages. Between those stages, the diffusion proceeds from initiated nodes. Thus, any top ranking nodes initiated by such diffusion is not used as seeds and in its place other high ranked but uninitiated node is chosen to be activated.

The diffusion model that is used here is stochastic. Even for single stage seeding, when the same set of seeds is initiated with repeated executions, the coverage will vary as each activated node will attempt to activate its neighbor with certain probability, but no certainty. Still, it can be proven that there broad sufficient conditions under which there exists a sequential seeding using the same node ranking method with average coverage exceeding that of the single stage seeding. One such sufficient but not necessary condition is that there is a seed in the single stage seeding that is reachable from other seeds by with probability smaller than 1. Indeed, by initiating all seeds except such reachable seed in the first stage, we can observe the following.  For all execution cases in which uninitiated seed is not activated by diffusion after the first stage, the average coverage over such cases will be the same for both methods. However, in all other execution cases (guaranteed to exist since the node is reachable with probability less than 1), the single stage seeding will have average coverage equal to the coverage of the first stage of the sequential seeding. However, in the second stage, the latter will initiate a new node in place of already active reachable seed (and perhaps more nodes in the subsequent diffusion), making the overall average coverage of the sequential seeding larger than the single stage one.

\section*{Results}

Sequential seeding strategies are considered here for an independent cascade model representing stochastic diffusion of information over the network initiated by seeds\cite{Kempe:2003}. A basic, commonly used, single stage ($SN$) seeding consists of only one stage with $n$ seeds. A seeding stage starts with activation of all $n$ seeds at the beginning to initiate diffusion and lasts until diffusion stops. According to the definition of independent cascade model each diffusion step consists of a single attempt performed by each node activated in the previous step to activate this node direct non-active neighbors with a given propagation probability. Due to the fact that we measure diffusion time in the number of steps, we assume that each such step takes a unit time to execute regardless of the network size. It means that time corresponds directly to the number of diffusion steps. Let $T_{SN}$ denote the number of diffusion steps of single stage seeding. All seeding methods considered here use the same fixed number of $n$ seeds. Five sequential seeding methods are introduced here as presented in Table ~\ref{tab:sq}; see the $Methods$ section for more details. In the single stage method, the $n$ seeds are utilized at once at the beginning, whereas new sequential approaches deploy them in a sequence of multiple consecutive stages. In general, there are two types of proposed sequential seeding. The first type uses constant seeding stage time and includes an approach denoted as SQ\_kPS with $k$ seeds initiated in each seeding stage. The other approach of this type denoted SQ\_TSN , uses the number of stages defined by the seeding time of single stage approach. The second type uses revival mode, in which the stage ends when the diffusion finishes and the next seeding stage starts immediately thereafter. This type includes approaches denoted as SQ\_kPS\_R, SQ\_TSN\_R and the revival mode with buffering denoted as SQ\_kPS\_B.

\begin {table}[!ht]
\begin{center}
\begin{tabular}{ l p{11cm} c c}
\hline
 & & \textbf{No. of} & \textbf{Steps}  \\ 
\textbf{Strategy} & \textbf{Description} & \textbf{seeding} & \textbf{per}\\
& & \textbf{stages} & \textbf{stage$^*$} \\ 
\hline
\textbf{SQ\_kPS}& \textit{k Seeds Per Stage} -- the sequential seeding consisting of $n$ stages; each activating $k$ additional seeds at its beginning and lasting one diffusion step & $n$ & $1$ \\ 
\hline
\textbf{SQ\_kPS\_R} & \textit{k Seeds Per Stage with Revival} -- with $k$ seeds invoking the stage and reviving the previously stopped diffusion; the stage ends when the diffusion stops & $n$ & $\geq 1$ \\ 
\hline
\textbf{SQ\_kPS\_B} & \textit{k Seeds Per Stage with Buffering} -- the seed set is distributed equally: $k$k seeds are assigned to each diffusion step which starts with $k$ seeds; the next seeding stage is applied after the diffusion started after the previous stage stops; the buffer collects for later use the unused seeds initiated by diffusion invoked by the previous seeding stages; the last seeding stage is in the nth diffusion step & $\leq n$ & $\leq n$ \\ 
\hline
\textbf{SQ\_TSN} & \textit{Time} from \textit{Single Stage} -- the number of steps $T_{SN}$ in the corresponding single stage seeding (SN) defines the number of seeding stages; seeds are distributed equally -- the number of seeds in each stage is the total number of seeds $n$ divided by the sequence length: ${n/T_{SN}}^{**}$ & $min(n,T_{SN})^{**}$  & $1$ \\ 
\hline
\textbf{SQ\_TSN\_R}  & \textit{Time} from \textit{Single Stage with Revival} -- the seeds are allocated among stages the same way as in SQ\_TSN, but a new stage starts and another set of ($n/T_{SN}$) seeds is initiated after the previously started diffusion stops $^{**}$ & $min(n,T_{SN})^{**}$ & $\geq 1$\\ 
\hline
\end{tabular}

\caption {New seeding strategies, all based on sequential approach either without or with revival mode; $n$ - denotes the fixed number of seeds, while $T_{SN}$ - the number of diffusion steps in the corresponding single stage approach (SN); $^*$ the number of steps in the last seeding stage is not defined and can be greater than 1, since the diffusion proceeds as long as possible. $^*{}^*$ if  $n<T_{SN}$, $k$ Seeds Per Stage (SQ\_kPS) approach with $n$ stages is used; otherwise seeding consists of $T_{SN}$ stages.}
\label{tab:sq} 
\end{center}
\end {table}

\begin{figure}[!ht]
\centering
\begin{minipage}{1\textwidth}
\centering
\includegraphics[width=\linewidth]{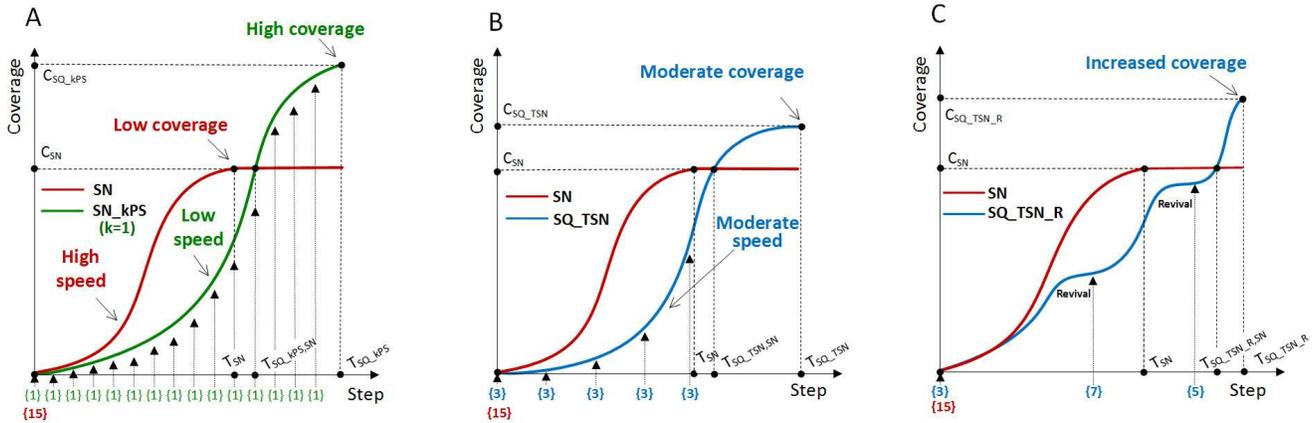}
\caption{\textbf{(A)} $k$ per stage sequential seeding strategy (SQ\_kPS) with $k=1$ compared  with the single stage approach (SN). \textbf{(B)}  Sequential strategy based on the reference time (SQ\_TSN) compared  with the single  stage approach (SN). \textbf{(C)}  Sequential strategy with revival mode (SQ\_TSN\_R) compared  with the single stage approach (SN).}
\label{fig:fig_1}
\end{minipage}\hfill
\end{figure}

The example of one seed per stage sequential seeding (SQ\_1PS) in the relation to single stage seeding (SN) with the same diffusion parameters is shown in Fig.~\ref{fig:fig_1}(A). The use of a single seed in each step of simulation instead of using all seeds at the beginning results in the increased coverage $C_{SQ\_kPS}$, when compared to single stage coverage $C_{SN}$. Time -- more precisely the number of diffusion steps -- when the maximal coverage is reached, denoted as $T_{SQ\_1PS}$ is also typically longer than the reference time for the single stage $T_{SN}$. Time $T_{SQ\_1PS,SN}$ denotes the number of steps when SQ\_1PS process reaches the maximal coverage $C_{SN}$ from single stage process. 

The second approach (SQ\_TSN) is illustrated in Fig.~\ref{fig:fig_1}(B), uses shorter sequence with the total number of stages equal to the number of steps needed to reach maximal coverage $C_{SN}$ in the single stage seeding (SN). The number of seeds $s$ used in each step is equal to the total number of available seeds $n$ divided by the number of steps $T_{SN}$ in the corresponding single stage process. An extension, based on the revival mode presented in Fig.~\ref{fig:fig_1}(C), uses additional seeds once the diffusion dies out and needs to be reseeded. The revival mode may also be extended with buffering: the number of seeds designated for initiation but not used because of their initiation in earlier diffusion steps is cumulatively stored in the buffer and the equivalent number of new not activated by diffusion nodes are initiated as seeds when the current stage diffusion terminates. 

Sequential seeding uses dynamic selection and any node ranking method including simple heuristics or their extensions can be applied. To show the advantages of sequential seeding over single stage approach, we use for both the same nodes ranking, and compare their performance for the following five rankings: random selection (R), degree (D), second level degree (D2) –- it takes into account node degree and the degrees of its neighbors, Page Rank (PR) and eigenvector (EV). In each seeding stage, only nodes that have been not activated yet are selected in the order of their initial ranking as seeds, making the sequential seeding set often different from the set used by the single stage approach. Rankings of nodes may be computed at the beginning of the process based on the structural measures within the initial network.

\begin{figure}[!ht]
\centering
\begin{minipage}{1\textwidth}
\centering
\includegraphics[width=\linewidth]{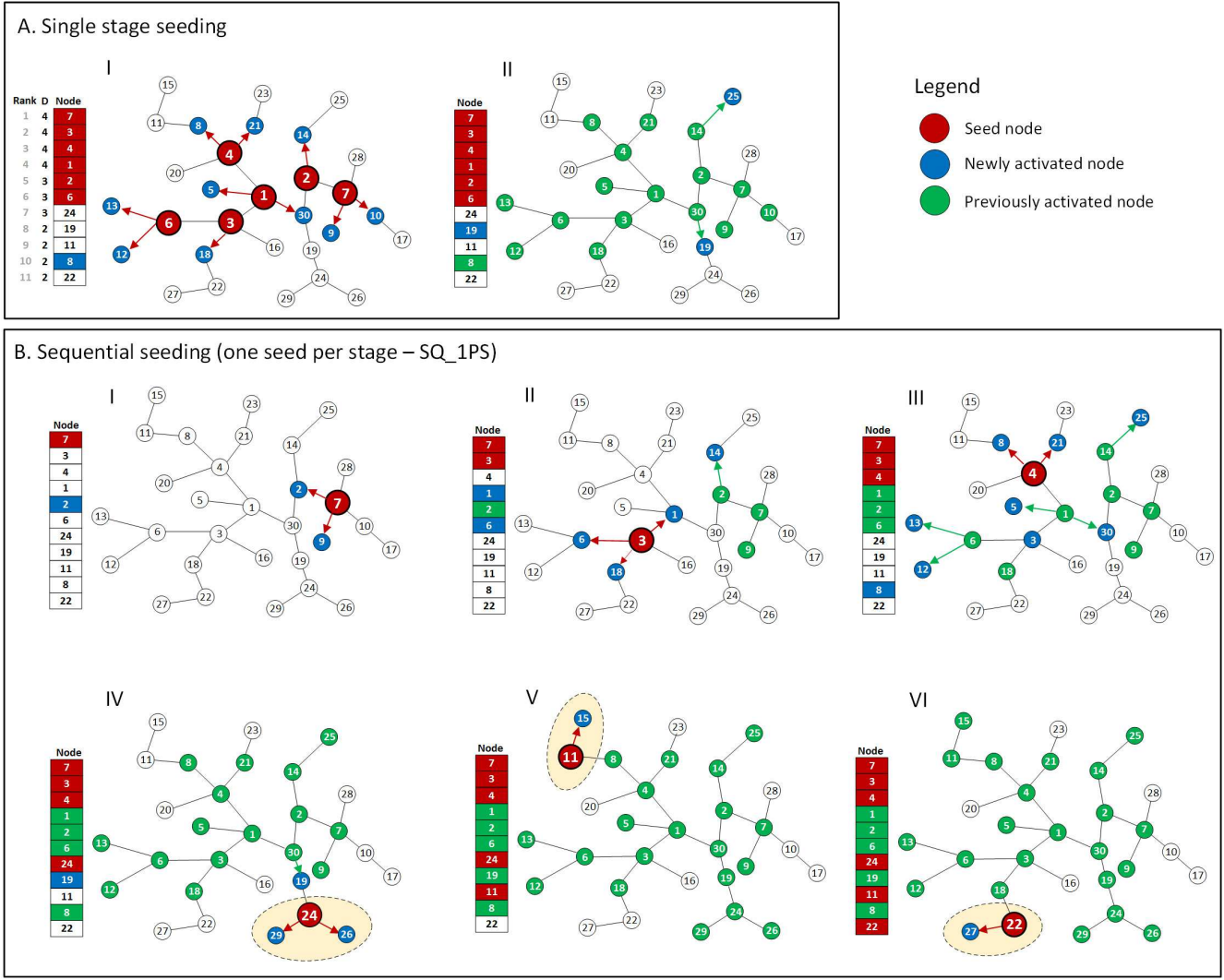}
\caption{\textbf{(A) Single stage seeding process (SN) with top degree ranking.} \textbf{(I)} An initial set of 6 seeds (red color) consists of the nodes with the highest degree ranking. The seeds activate with probability PP=0.5 their neighbors causing 10 of them to become active. In the next step \textbf{(II)}, all newly activated nodes are contacting their non-active neighbors, activating two of them. This process ends with the total of 18 activated nodes. \textbf{(B) One per stage sequential seeding strategy (SQ\_1PS) with top degree node ranking.} \textbf{(I)} A node with the highest degree is selected as the first seed. Since four nodes have the same degree, node 7 is selected randomly. It tries to activate its neighbors with PP=0.5 and nodes 2 and 9 become active. Node 2, seeded in SN (AI) is activated by diffusion. \textbf{(II)} Node 3 is the second seed. At this stage, three nodes being seeds in the SN process, i.e. {1,2,6} are activated by other nodes; three other nodes can take their place as seeds. \textbf{(III)} Node 4 is taken as the third seed. \textbf{(IV)} The fourth seed is node 24 -- it did not belong to seeds in the SN process. It could become a seed because all nodes preceding it in the ranking have already been activated. So far, 20 nodes are activated in four steps comparing to 18 in the SN strategy, and there are two more seeds left to activate. \textbf{(V)} The fifth seed is node 11. Since the other nodes do not activate any new ones, the process would terminate now with totally 22 activated nodes. However, one more seed is still available. \textbf{(VI)} The last seed is node 22 which activates node 27. It cannot activate anyone else, so diffusion stops with 24 activated nodes. New parts of the network reached with the sequential seeding are marked.}
\label{fig:fig_3}
\end{minipage}\hfill
\end{figure}

The results achieved with the sequential seeding strategies are compared, Table ~\ref{tab:par}, with the single stage seeding (SN) for 1,875 versatile configurations. For each of the 15 networks, we use the same diffusion parameters (five propagation probabilities), the total number of seeds (five values of SP) and five node rankings. The simulation outcomes from all sequential strategies compared with single stage approach manifests the improvement in 89.8\% configurations, with the largest benefit, in terms of better coverage, of 10.1\% on average but exceeding 50\% in some cases. Simultaneously, sequential methods increase the diffusion duration by 5.2 times on average. It is worth noticing that the average coverage achieved with sequential seeding SQ\_1PS\_R applied to any seed ranking method is greater than the average coverage provided by the single seeding based on the same ranking. 
We observed two main factors affecting the performance of sequential seeding. The first is the extent to which sequential seeding avoids selecting nodes that would be activated through diffusion. The gain arises when a top ranked but already activated node is removed from  selection and in its place another not yet activated node becomes a seed,  Fig.~\ref{fig:fig_3}(B), ), stages IV--VI. It is most likely to happen in happen in the revival mode, which maximally delays the next seeding stage until diffusion stops in the current stage.
 

The performance improvement depends also on selecting new seeds that have high potential of activating their neighbors in place of those seeds that were already activated by diffusion. The gain from sequential approach is commonly greater for configurations with low overall coverage, i.e. low propagation probability, see  Fig.~\ref{fig:fig_2}(D2), networks N1, N4, N5, N7 described in Supplementary Information, Fig.~\ref{fig:fig_2}(D4) as well as for large seed sets, see Fig.~\ref{fig:fig_2}(D3).

Distributing the seeds over time makes it possible to take advantage of the progress the diffusion made before new seeds are selected. It is important especially when the budget, i.e. the number of initial seeds, is fixed. In such a case under single seeding approach the entire budget of seed is consumed at the outset and recovery is not possible if diffusion stops without reaching the desired coverage. Gradual spending allows to fund a campaign for a longer time period and to reach more customers. 

The resolution of trade-off between the final coverage and speed of diffusion depends on individual preferences for the importance of both coverage and the process duration in specific applications. Better coverage can be achieved by sequential approach, but it takes more time to finish or even to reach the same coverage as for single stage seeding. The more stages the sequential seeding process has, the more nodes might be activated, but the longer diffusion lasts. If the process is optimized for time, the length of sequence should be short. For example during the elections, time is fixed and all possible efforts should be taken to acquire as many voters as possible by the voting date. With high speed epidemics, vaccination should be applied quickly to protect as many people as possible. In critical situations, when initial actions are not sufficient, additional resources can be engaged and additional seeds activated. 

An experimental setup runs agent-based simulations on 15 static real networks N1-N15, containing from 899 to 16,726 nodes and from 2,742 to 147,547 edges, see Supplementary Information for details.

The independent cascades model (IC)\cite{Kempe:2003} was used with propagation probability $PP(a,b)$ that node $a$ activates (influences or infects) node $b$ in the step $t+1$ under condition that node $a$ was activated at time $t$\cite{Wang:001}. The main reason for selecting this model was a relatively small number of seeds needed in each stage for the optimal decomposition of initial set of seeds. With this model even a single seed can induce diffusion, while in linear threshold model\cite{Kempe:2003} (LT) a small seeds packages would often have no effect. Parameters used in simulations characterizing diffusion, networks and strategies are presented in Table ~\ref{tab:par}.

\begin {table}[!ht]
\begin{center}
\begin{tabular}{c l c p{7.5cm} } 
\hline
\textbf{Symbol}& \textbf{Parameter}& \textbf {No. of distinct values} & \textbf {Variants} \\ 
\hline
N & Network & 15 &Real networks from various areas\\ 
\hline
PP & Propagation probability& 5&0.05, 0.1, 0.15, 0.20, 0.25 \\ 
\hline
SP & Seeds percentage & 5&1\%, 2\%, 3\%, 4\%, 5\% \\ 
\hline
S & Node ranking &5&random (R), degree (D), second level degree (D2),  \\ 
& for seed selection& &  Page Rank (PR), eigenvector (EV)\\
\hline
\end{tabular}
\caption {Networks and parameters of diffusion used in simulations}
\label{tab:par} 
\end{center}
\end {table}

Simulation parameters create an experimental space N 
$\varprod$ PP $\varprod$ SP $\varprod$ S, resulting in 1,875 configurations. Each configuration was independently applied for each seeding strategy from Table ~\ref{tab:sq} and the results were averaged over 100 runs for each case. The agent based model was used, with agents connected according to the networks specifications. Each step of simulation includes the selection of additional seeds (if additional seeding is allowed), visiting all nodes activated in the previous step and newly selected seeds, and finally activating their neighbors according to the propagation probability (PP). Within the experimental space sequential seeding algorithms were tested including constant seeding stage time and revival mode, see Table  ~\ref{tab:sq}. 

The simulation results achieved in sequential seeding were compared with the single stage seeding (SN) on the same network (N) and with the same parameters including propagation probability (PP), seeding percentage (SP) and node ranking strategy (S). Reference values for comparison were based on the coverage achieved in the single stage seeding $C_{SN}$ with the duration representing the number of steps in the single stage process $T_{SN}$ when the coverage $C_{SN}$ is achieved, see Fig.~\ref{fig:fig_1} and the $Methods$ section.

During experiments several variants of sequential seeding approach were verified including k seeds per stage with  $k$ equal to 1, 2, 4 and 8 with both revival (SQ\_kPS\_R) and non-revival mode (SQ\_kPS). For $k=1$ buffered approach was used (SQ\_1PS\_B). Reference based strategy was verified with revival (SQ\_TSN\_R) and non-revival mode (SQ\_TSN). Finally 11 variants of sequential strategies were used.

The best results in terms of coverage were delivered by the sequential strategy with $k=1$ seed used per stage with revival mode (SQ\_1PS\_R) but they required the largest number of steps to finish. The opposite results with the shortest duration and the lowest coverage were achieved by the sequential approach with reference time from single stage  (SQ\_TSN), see Fig.~\ref{fig:fig_5}(D) for the results from both strategies. 

The SQ\_1PS\_R method achieved the highest decomposition of the seeding and the highest usage of single seeds which resulted in the highest coverage, but it also have diffusion taking the longest time to finish. For SQ\_TSN, the number of seeding stages and the coverage ware usually low. Both strategies were compared with the coverage of single stage strategy (SN) achieved for the same simulation parameters, Fig.~\ref{fig:fig_2}(A). The results from 1,875 configurations for both strategies are ordered by coverage obtained in the single stage method represented by red line. For the parameters with relatively low coverage, below 5\%, results for the first 250 cases for all strategies are quite similar since the differences are small, while with growing coverage the differences are much more visible. Sequential seeding is almost always better than its single stage equivalent with the same parameters. The global results for all networks, strategies and parameters show better results than SN in 95.3\% of simulation cases delivered by SQ\_1PS\_R. The gain was also observed in 85.49\% cases for SQ\_TSN. Fig.~\ref{fig:fig_2}(B) illustrates coverage performance of sequential seeding strategies, SQ\_1PS\_R and SQ\_TSN represented as a ratio of coverage obtained by these two methods to the one achieved the single stage (SN). The length of sequence positively affects the total results. In 1,697 cases from 1,875 (90.5\%), one seed per stage method (SQ\_1PS\_R) outperformed seeding with shorter sequences (SQ\_TSN). The example comparison of simulation results between SQ\_TSN, SQ\_1PS\_R and single stage seeding is presented in Fig.~\ref{fig:fig_2}(C). Sequential strategies outperform the single stage approach even though they need more time (steps) to achieve that.

\begin{figure}[!ht]
\centering
\begin{minipage}{1\textwidth}
\centering
\includegraphics[width=\linewidth]{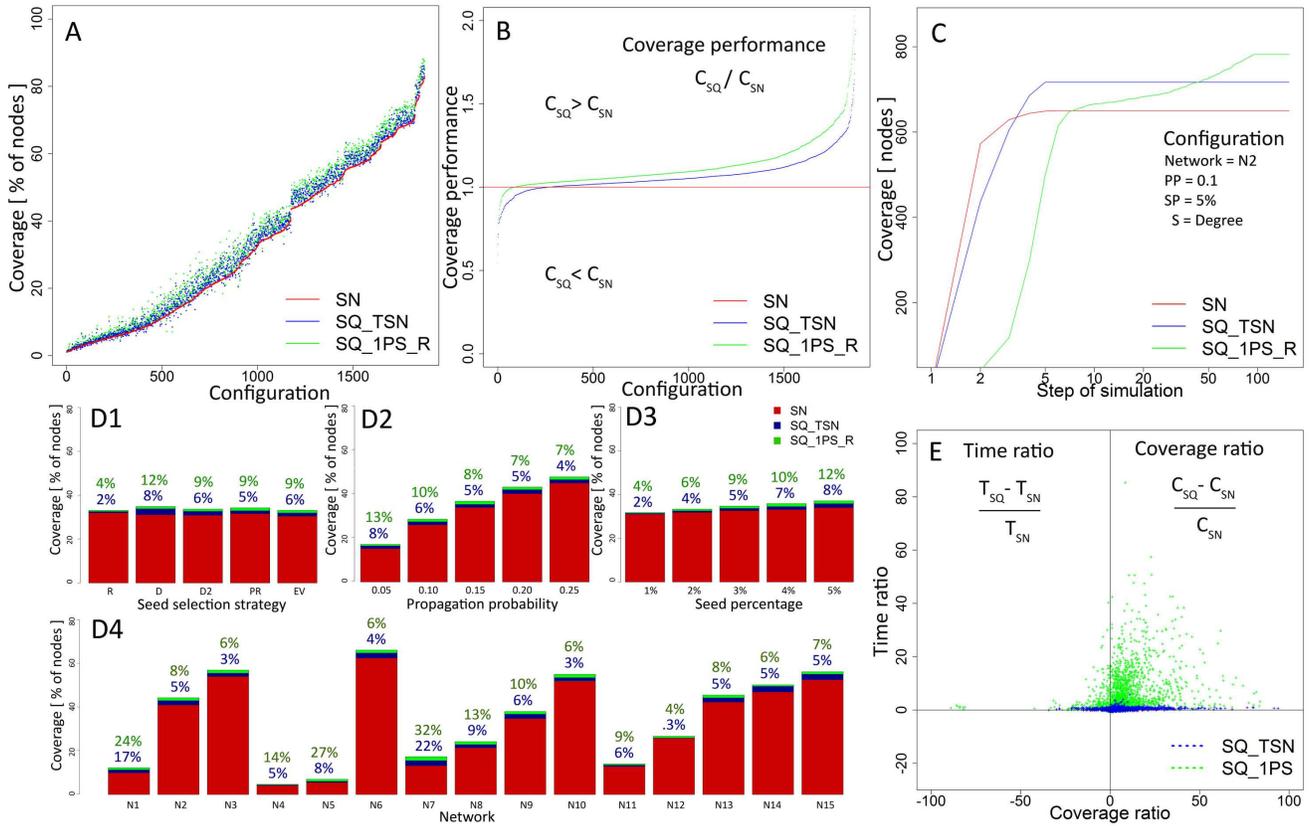}
\caption{ \textbf{(A)} Average coverage from all 1,875 configurations for $k=1$ per stage seeding  (SQ\_1PS\_R) and the sequential approach based on single stage time reference (SQ\_TSN), compared with the single stage method (SN) for all simulation cases ordered by single stage coverage; \textbf{(B)} Performance of $k=1$ per stage seeding  (SQ\_1PS\_R) and reference time based seeding (SQ\_TSN) in the relation to single stage (SN) for all configurations; \textbf{(C)} Cumulative coverage results from 100 steps of simulation with degree based ranking (D) for network N2 with propagation probability PP=0.1 and seeding percentage SP=5\%; \textbf{(D1)} Coverage increase for SQ\_1PS\_R and SQ\_TSN based on various seed ranking methods: random R, degree D, second level degree D2, Page Rank PR, Eigenvector EV; red part shows single stage seeding (SN) coverage, blue part and font show sequential SQ\_TSN improvement while green part and font show sequential SQ\_1PS\_R gain; \textbf{(D2)} Coverage increase for SQ\_1PS\_R and SQ\_TSN for all used propagation probabilities with values 0.05, 0.1, 0.15, 0.20 and 0.25; \textbf{(D3)} Results for seed selection percentages from the range 1\% to 5\% for SQ\_1PS\_R and SQ\_TSN; \textbf{(D4)} Average coverage from single stage seeding together with coverage increase from sequential seeding SQ\_1PS\_R and SQ\_TSN for networks N1 - N15; \textbf{(E)} Distribution of simulation cases in terms of coverage and time ratios for $k=1$ per stage seeding (SQ\_1PS\_R) and reference time based seeding (SQ\_TSN) for all networks, where $T_{SQ}$ and $C_{SQ}$  denote time and coverage in appropriate sequential seeding: SQ\_1PS\_R or SQ\_TSN, while $T_{SN}$ and $C_{SN}$ stand for time and coverage in single stage seeding}
\label{fig:fig_2}
\end{minipage}\hfill
\end{figure}

The results from simulations demonstrate that the improvement can even exceed 50\% with the use of the same number of seeds as in the single stage seeding. The coverage of diffusion in SQ\_1PS\_R is on average 13.1\% better than when using SN approach with the same parameters. The SQ\_TSN increased the coverage, on average by 7.15\%, comparing to the single stage seeding. 

The performance of proposed methods is dependent on parameters of diffusion and network characteristics. The results are also dependent on seed ranking strategy used. For example, the coverage gain and its distance from the results of the single stage seeding for both SQ\_1PS\_R and SQ\_TSN was the highest for degree based ranking and the lowest for random node ranking as seen Fig.~\ref{fig:fig_2}(D1). The overall coverage level depends only little on strategy or seed percentage and it is much more impacted by propagation probability and network profile. Experiments were carried out on a wide range of networks and parameters, including cases with very low performance/coverage, e.g. the propagation probability PP=0.05 or seeding percentage SP=1\%. 

For such parameters, it is very difficult to trigger a cascade regardless of seeding strategy used. The highest improvement was observed for the lowest propagation probability PP=0.05 at 13\% for SQ\_1PS\_R compared to single stage seeding with SQ\_TSN delivering 8\% improvement. With the highest propagation probability PP=0.25 the improvement decreases to 7\% and 4\% for both strategies respectively, see Fig.~\ref{fig:fig_2}(D2). In configurations with high propagation probability, diffusion reaches high coverage in a very short time, regardless of what seeding strategy is used - there is almost no space for any improvement. In terms of seeding percentage (SP) performance of sequential seeding grows together with the number of seeds selected. The lowest seeding percentage SP=1\% delivered 4\% and 2\% improvement while for the SP=5\% the improvement increases to 12\% and 8\% respectively, Fig.~\ref{fig:fig_2}(D3). In general, the obtained results were strongly dependent on the network profile with the overall coverage and performance of sequential seeding illustrated in Fig.~\ref{fig:fig_2}(D3) and in Fig.~\ref{fig:fig_5}(A).

One of the reasons for a generally good coverage performance of sequential methods, is related to identifying new seeds with the highest potential to activate other nodes. Sequential seeding efficiently utilizes the potential of diffusion, but spreading seeds over time increases the diffusion duration, see Fig.~\ref{fig:fig_2}(C). The greatest coverage can be achieved by means of one per stage strategy with revival SQ\_1PS\_R, but it results in the longest duration since every single seed waits until diffusion started by the previous seed terminates. On the other hand, the coverage performance of the SQ\_TSN strategy is lower, but it takes less time to reach the coverage greater than in the single stage approach, Fig.~\ref{fig:fig_2}(C).

The final decision which strategy to use depends on how the trade-off between time and coverage is resolved. If the process is restricted in time, the seeding sequence length should be shorter. For SQ\_TSN, the diffusion is on average only 1.4 times longer than for the single stage approach SN whereas the duration of the longest sequences based on SQ\_1PS\_R is on average 11.9 times longer. 

Gain in coverage (positive coverage ratio) and loss in time (positive time ratio) for both sequential strategies seen in Fig.~\ref{fig:fig_2}(E) presents a trade-off between acceptable extended time and desired increased coverage for two sequential methods SQ\_1PS\_R and SQ\_TSN shown in Fig.~\ref{fig:fig_2}(E). It can be seen that SQ\_1PS\_R provides greater variety of possible solutions and much greater gain in coverage, at the cost of the substantial expense of speed. 

While the results from two opposite approaches SQ\_1PS\_R and SQ\_TSN are presented in this section, their extensions based on revival mode and buffering are described in the $Discussion$ section. The SQ\_kPS strategy and usage of $k$ parameter other than one (2, 4, 8) makes it possible to accordingly adjust both duration and the coverage.  

\section*{Methods}

The common approach for seeding in complex networks uses a single stage seeding (SN), Fig.~\ref{fig:fig_3}(A). In the first stage, all $n$ initial seeds are selected using a given node ranking. In the next and last stage, diffusion starts, and then continues without any additional support until it stops at certain time to which we refer as saturation time $T_{SN}$, achieving  coverage $C_{SN}$. An sample network with 30 nodes was generated using Baraba´si-Albert model. At the beginning, $n=6$ nodes with the highest degrees, representing 20\% of all nodes, are selected as seeds. Then diffusion is simulated using the Independent Cascades model with propagation probability P=0.5, see  Fig.~\ref{fig:fig_3}(AI) and (AII).

In the sequential seeding approach, introduced here, the seeds are split among several seeding stages interspersed with diffusion stages so each seeding stage is followed by diffusion stage. The selection of seeds in each stage is based on inactive node ranks. The first sequential method –- one seed per stage  (SQ\_1PS) utilizes the highest distribution of seeds assigning a single seed to each seeding stage. The number of stages equals the number of seeds - $n$. To illustrate this process the same network with the same number of seeds as for SN is used in Fig.~\ref{fig:fig_3}(B). The seeding consists of six stages with one seed per stage. In each stage, a new seed with the highest degree is selected from not activated yet nodes. It means that the already activated nodes are not considered for selection, even if they possess the high rank. This allows avoiding to seed nodes that would be activated anyway by diffusion. 

The diffusion with sequential seeding (SQ\_1PS) reaches 24 nodes, while the single stage seeding (SN) activates just 18. All those nodes, except node 11 that have been activated by the SN method, were also activated by sequential strategy which in addition activated seven more nodes. However, since diffusion is simulated using the probabilistic Independent Cascades model, the outcome of sequential seeding strategy might be different, or even worse, than the final result of single stage seeding strategy. Nevertheless, as proved by experimental results, the sequential seeding strategy outperforms the single stage method in over 90\% of cases.

\begin{figure}[!ht]
\centering
\begin{minipage}{1\textwidth}
\centering
\includegraphics[width=\linewidth]{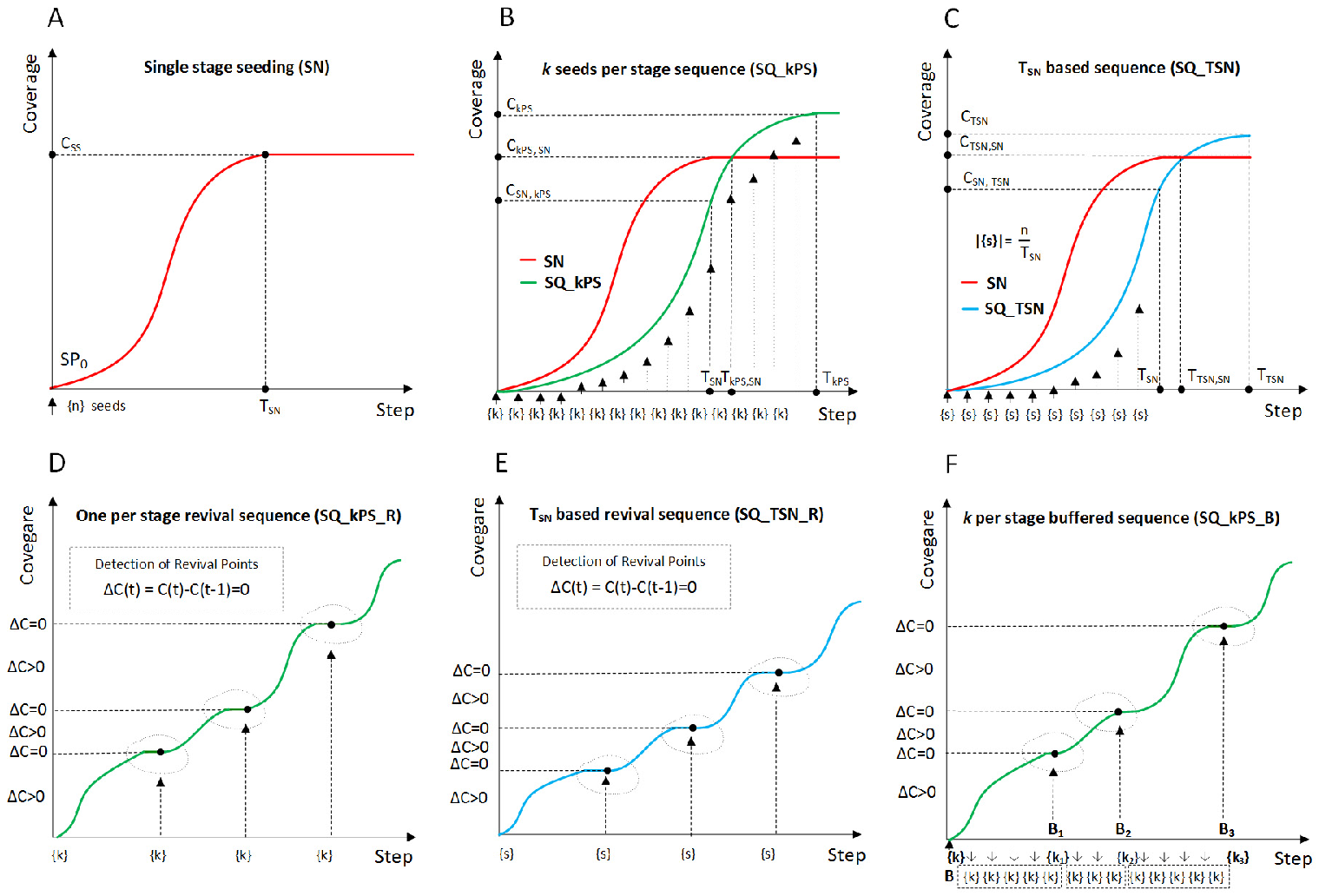}
\caption{\textbf{(A)} Diffusion based on the single stage seeding. 
\textbf{(B)} $k$ per stage sequential seeding strategy SQ\_kPS.  
\textbf{(C)} Sequential seeding based on reference time T$_{SN}$ -- SQ\_TSN.  
\textbf{(D)} $k$ per stage sequential seeding strategy with the revival mode SQ\_kPS\_R.
\textbf{(E)} The T$_{SN}$-based sequential seeding with the revival mode SQ\_TSN\_R.
\textbf{(F)} $k$ per stage sequential seeding with buffering SQ\_kPS\_B.}
\label{fig:fig_4}
\end{minipage}\hfill
\end{figure}

The higher coverage and the longer duration compared to single stage SN shown in Fig. ~\ref{fig:fig_4}(A) is illustrated for the $k$ per stage sequential seeding process SQ\_1PS in Fig.~\ref{fig:fig_4}(B). The questions arise what is the increased coverage $C_{OPS}$ how much it is greater than $C_{SN}$ and how much longer is the time $T_{OPS}$ to achieve this coverage greater than $T_{SN}$. Additionally, we can assess the lower speed of the SQ\_1PS process by measuring the time $T_{OPS,SN}$ when the maximal coverage $C_{SN}$ from the SN process is reached. Commonly, $T_{OPS,SN}$ \textgreater $T_{SN}$. Another sign of lower speed is that the coverage $C_{SN,1PS}$ of the SQ\_kPS process is achieved in time $T_{SN}$ and typically, $C_{SN,kPS}$ \textless $C_{SN}$.

While the sequences with the SQ\_1PS are relatively long, the second proposed approach is based on the reference time $T_{SN}$ from the single stage seeding applied to sequential approach as SQ\_TSN, Fig.~\ref{fig:fig_4}(C). The number of seeds used in each step is equal to the number of seeds $s$ used in single stage seeding divided by the number of steps $T_{SN}$ needed to reach the saturation point in SN. The sequence of seeding starts at the first stage and finishes at $T_{SN}^{th}$ stage. The coverage $C_{TSN,SN}$ of the SQ\_TSN process at the time $T_{SN}$ can be compared with $C_{SN}$. The slower speed of the SQ\_TSN means that $C_{TSN,SN}$ \textless $C_{SN}$. However, $C_{TSN,SN}$ \textgreater $C_{OPS,SN}$, i.e. SQ\_TSN is faster than SQ\_kPS. It can be observed that $T_{OPS,SN}$ \textgreater $T_{TSN,SN}$ \textgreater $T_{SN}$. On the other hand, regarding the total coverages, we have: $C_{OPS}$ \textgreater $C_{TSN}$ \textgreater $C_{SN}$. Overall, the greater coverage is, the lower speed is. 

Sequential seeding with revival and buffering modes. SQ\_kPS and SQ\_TSN are able to improve coverage thanks to the allocation of additional seeds, even if diffusion is still ongoing. To utilize the acquired knowledge about process dynamics, the sequential strategy with revival mode is proposed and applied to modify sequential strategies with constant seeding stage time: SQ\_kPS\_R and SQ\_TSN\_R. In both cases, an additional seeding is suspended until the most recent diffusion stops, i.e., the increase of coverage in two following steps $\Delta$C is zero, see Fig.~\ref{fig:fig_4}(D) and (E). It means that the proposed approaches extend the execution time of individual seeding stages and the total process duration. Only one seed per stage is used in SQ\_kPS\_R and similarly to SQ\_TSN, the allocated package of $|s|=n/T_{SN}$ seeds is used in SQ\_TSN\_R. According to the nature of the independent cascades model, each recently activated node has only one chance to infect its neighbors. Hence, if the process stops, such a node would never be re-activated in the future stages. The only way to restart diffusion is by initiating additional seed (SQ\_kPS\_R) or seeds (SQ\_TSN\_R). This can improve the final diffusion coverage, but at the price of spread increased duration.

Yet another proposed approach SQ\_kPS\_B is based on one seed per stage method and accumulates in a buffer the number of seeds unused because they were initiated by diffusion. The equal number of inactive nodes is initiated after diffusion terminates and requires recovery. The main goal of SQ\_kPS\_B is to limit the process duration and make it equal to SQ\_kPS, duration, simultaneously using the potential of diffusion, Fig.~\ref{fig:fig_4}(F). If we compare SQ\_kPS\_B with SQ\_kPS, the diffusion time of both is similar but the coverage for the buffering mode -- SQ\_kPS\_B is commonly greater because of the better exploitation of diffusion. 

\begin {table}[!ht]
\begin{center}
\begin{tabular}{l c c c c c c} 

\hline
 & \multicolumn{3}{c}{\textbf{Sequential compared to single stage SN}} & \multicolumn{3}{c}{\textbf{SQ\_1PS\_R compared to other sequential}} \\ 
\hline
\multicolumn{1}{c}{\textbf{SQ}} & \textbf{Cases}& \textbf {Coverage} & \textbf {Duration} & \textbf{Cases} & \textbf {Coverage} & \textbf {Duration}   \\ 
\multicolumn{1}{c}{\textbf{Method}} & \textbf{with}& \textbf {increase} & \textbf {increase} & \textbf{with} & \textbf {increase} & \textbf {increase}  \\ 
 & \textbf{$C_{SQ}>C_{SN}$ }& \textbf {$C_{SQ}$ vs. $C_{SN}$} & \textbf {$T_{SQ}/T_{SN}$} & \textbf{$C_{SQ\_1PS\_R}$} & \textbf {$C_{SQ\_1PS\_R}$} & \textbf {$T_{SQ\_1PS\_R}/T_{SQ}$}  \\ 
 & &  &  & \textbf{$>C_{SQ}$ }& \textbf {vs. $C_{SQ}$} &   \\

\hline
SQ\_1PS & 85.8\% & 8.8\% \ ($\Delta$=67.6) & 7.3 ($\Delta$=102) & 82.5\% & 6.2\% ($\Delta$=14.2) & 1.7 \ ($\Delta$=66)\\ 
SQ\_2PS & 86.9\% & 9.0\% \ ($\Delta$=66.1)& 3.9 ($\Delta$=46.5) & 85.3\% & 3.8\% ($\Delta$=17.2) & 2.9 ($\Delta$=126)\\ 
SQ\_4PS & 84.3\% & 8.0\% \ ($\Delta$=61.3) & 2.3 \ \ ($\Delta$=20) & 87.9\% & 4.8\% ($\Delta$=22.8) & 4.7 ($\Delta$=155)\\ 
SQ\_8PS & 82.2\% & 7.0\% \ ($\Delta$=53.9) & 1.6 \ \ ($\Delta$=8.0) & 89.0\% & 5.9\% ($\Delta$=30.5) & 6.7 ($\Delta$=168)\\ 
SQ\_1PS\_R & 95.3\% & 13.1\% ($\Delta$=85.5) & 11.9 ($\Delta$=176) & -- & -- & -- \\ 
SQ\_2PS\_R & 95.1\% & 12.9\% ($\Delta$=85.9) & 7.7 ($\Delta$=107) & 51.4\% & 0.2\% \ ($\Delta$=0.5) & 1.5 \ ($\Delta$=68)\\ 
SQ\_4PS\_R & 94.2\% & 12.1\% ($\Delta$=82.2) & 5.2 ($\Delta$=65.5) & 59.0\% & 0.9\% \ ($\Delta$=3.2) & 2.1 ($\Delta$=108)\\ 
SQ\_8PS\_R & 93.1\% & 11.1\% ($\Delta$=79.2) & 3.6 ($\Delta$=40.5) & 66.8\% & 2.1\% \ ($\Delta$=7.4) & 2.9 ($\Delta$=134)\\ 
SQ\_TSN & 85.5\% & 7.2\% \ ($\Delta$=54.1) & 1.4 \ ($\Delta$=7.5) & 90.5\% & 5.5\% ($\Delta$=31.5) & 8.8 ($\Delta$=169)\\ 
SQ\_TSN\_R & 94.2\% & 11.4\% ($\Delta$=77.7) & 3.2 \ ($\Delta$=42) & 65.5\% & 1.8\% \  ($\Delta$=8.2) & 3.5 ($\Delta$=134)\\ 
SQ\_1PS\_B & 91.3\% & 10.7\% ($\Delta$=76.1) & 9.3 ($\Delta$=133) & 63.2\% & 2.2\% \  ($\Delta$=7.0) & 1.3 \ ($\Delta$=35)\\ 
\hline
\textbf{Average} & \textbf{89.9\%} & \textbf{10.1\% \ \ \ \ \ \ \ \ \ \ \ \ \ \ \ \ } & \textbf{5.2\ \ \ \ \ \ \ \ \ \ \ \ \ \ \ \ } & \textbf{74.1\%} & \textbf{3.3\%\ \ \ \ \ \ \ \ \ \ \ \ \ \ \ \ } & \textbf{3.6\ \ \ \ \ \ \ \ \ \ \ \ \ \ \ } \\ 
\hline

\end{tabular}
\caption {Coverage and duration of various sequential methods compared to the worse single stage seeding SN and the best sequential approach SQ\_1PS\_R; $\Delta$ - Hodges-Lehmann estimator used in Wilcoxon signed rank test as a measure of the difference between two result groups; values $\Delta>0$ demonstrate significantly higher values for coverage and duration of all sequential strategies compared to single stage approach SN and SQ\_1PS\_R compared to other sequential approaches; all analysis achieved statistical significance p-value \textless 2e-16.}
\label{tab:t4} 
\end{center}
\end {table}

\section*{Discussion}

Sequential seeding with one seed per stage SQ\_1PS outperforms coverage of the single stage seeding SN in over 85\% of configurations, providing on average nearly 9\% better coverage but its duration was over 7 times longer, Table ~\ref{tab:t4}. 

\begin{figure}[!ht]
\centering
\includegraphics[width=500px]{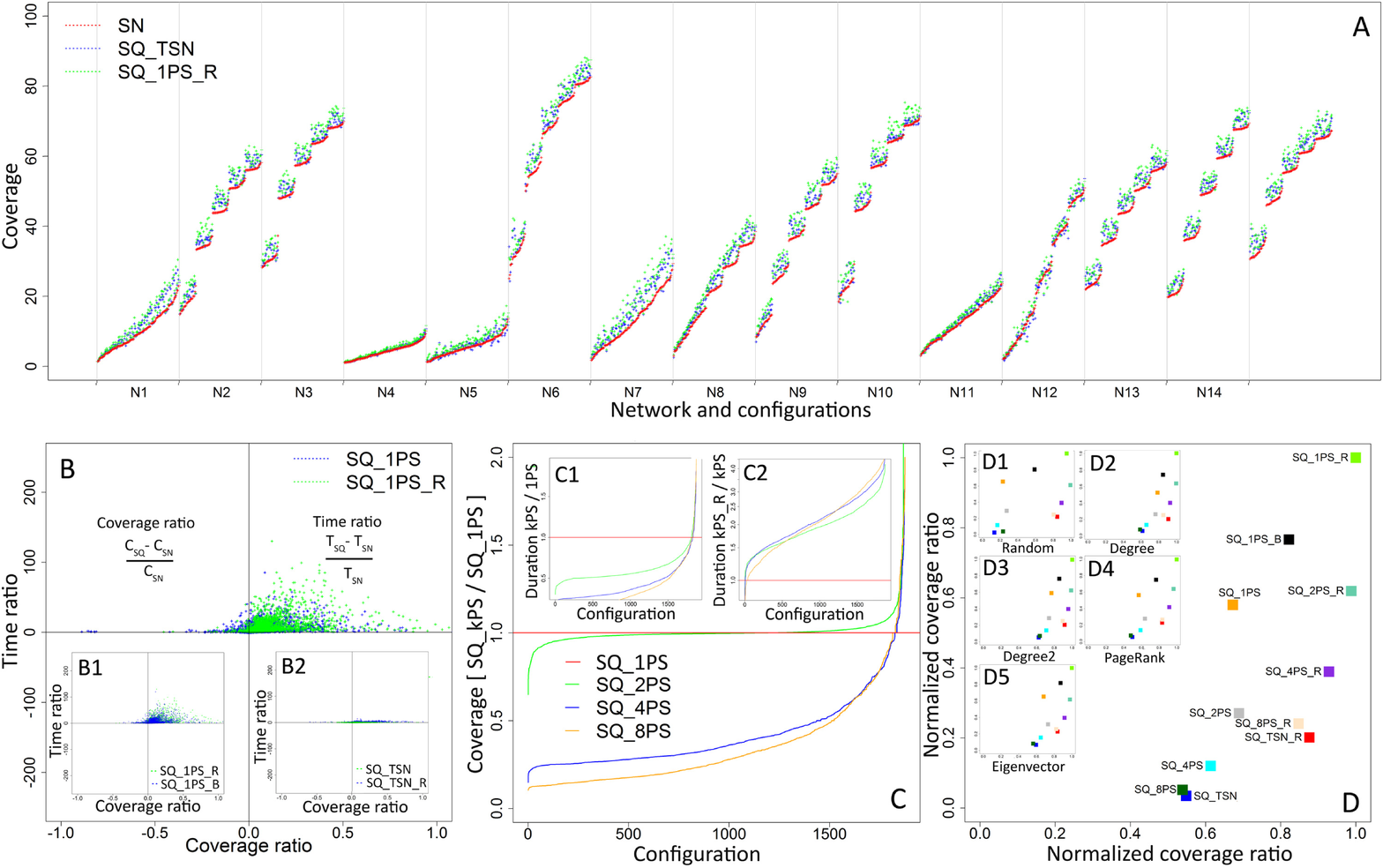}
\caption{\textbf{(A)} Coverage for SQ\_1PS\_R and SQ\_TSN compared with single stage (SN) individually for each network;  \textbf{(B)} Speed and coverage ratios for one per stage strategy with(SQ\_1PS\_R) and without revival mode (SQ\_1PS) in reference to single stage approach SN; $T_{SQ}$, $C_{SQ}$ -- time and coverage in suitable sequential seeding;\textbf{(B1)} Time and coverage ratios for sequential strategy with revival SQ\_1PS\_R vs. sequential strategy with buffering SQ\_1PS\_B; \textbf{(B2)} Reference based sequential strategy SQ\_TSN vs. SQ\_TSN\_B ; \textbf{(C)} Coverage ratios for SQ\_kPS strategies with 2, 4 and 8 seeds per stage in relation to one per stage approach SQ\_1PS; \textbf{(C1)} Duration of diffusion for SQ\_kPS strategies with 2, 4 and 8 seeds per stage in relation to one per stage approach SQ\_1PS; \textbf{(C2)} Duration of SQ\_kPS\_R strategies with revival in the relation to SQ\_kPS without revival; \textbf{(D)} Normalized average time and coverage ratios for all used sequential strategies; \textbf{(D1)} Normalized ratios for random (R) node ranking; \textbf{(D2)} Normalized ratios for degree based (D) node ranking; \textbf{(D3)} Normalized ratios for second level degree (D2) based node ranking; \textbf{(D4)} Normalized ratios for PageRank (PR) based node ranking; \textbf{(D5)} Normalized ratios for eigenvector based node ranking.}
\label{fig:fig_5}
\end{figure}

The revival mode applied to single seed per stage (SQ\_1PS\_R) achieves the best coverage among all sequential methods - better by over 13\% on average with gained observed for more than 95\% of configurations but lasts the longest (over 10 times longer than SN). In general, the revival mode applied to any $k$ per stage methods (SQ\_kPS) outperforms single stage approach in 10\% more configurations with 43\%-59\% increase of coverage gain compared to their non-revival version, see also Fig.~\ref{fig:fig_5}(B). The similar improvement is observed for application of revival mode to SQ\_TSN, Fig.~\ref{fig:fig_5}(B2). The revival mode provides greater coverage than its reference version since it makes better use of interstage diffusion. On the other hand, the revival mode significantly extends the process duration: by 60\% to 130\%. 

The buffered mode SQ\_1PS\_B gains little compared to the regular one per stage method SQ\_1PS at the small expense of time so it provides limited SQ\_1PS improvement, much weaker than provided by the revival mode, Fig.~\ref{fig:fig_5}(B1).

The experiments include some configurations with very low final coverage, which result from low propagation probability PP=0.05 combined with low seeding percentage SP=1\%. It means that the total number of activations will always be low, no matter what strategy is used. For example, for a small network N10 with SP=1\% only 9 seeds are activated, so it is very difficult to induce any diffusion with PP=0.05. An opposite case arises for networks with very high node degrees and relatively high propagation probability, i.e. PP=0.25 and SP=5\%, where all strategies perform very well resulting in a high number of activated nodes in a very short time. Taking into account the above conditions, an increase by 13\% on average for SQ\_1PS\_R can be considered substantial. 

Among all sequential strategies SQ\_TSN is the fastest -- it increases duration in reference to SN by only 1.4 times, but its coverage gain is relatively little: 7\%. Hence, two strategies: SQ\_1PS\_R and SQ\_TSN can be treated as upper and lower sequential boundaries, respectively, both for coverage and time, Fig.~\ref{fig:fig_5}(D). The space between them may be explored by using different values of seeds per stage $k$ either with or without revival: SQ\_kPS\_R or SQ\_kPS for $k=2, 4, 8, etc$.  

Overall, the experimental results reveal that the proper seeding strategy is a trade-off between time and coverage, which can be resolved according to the user needs by choosing appropriate sequential seeding strategy. Fig.~\ref{fig:fig_5} and  Table ~\ref{tab:t4} provide details on performance of the introduced sequential seeding strategies in terms of trade-off values. The greatest coverage can be achieved by means of longer sequences resulting in longer execution. If the process is limited in time that is very short, the sequential strategies may be less justified. 

Here we showed both experimentally and theoretically that the average coverage of any strategy of ranking nodes for selection of all seeds at the beginning of the process can be improved with sequential seeding using the same ranking. This is because the proposed method is dynamic as it adapts the set of activated seeds to the progress of the execution of diffusion active between stages of seeding.

\vspace{\baselineskip}
\vspace{\baselineskip}
\vspace{\baselineskip}
\vspace{\baselineskip}
\vspace{\baselineskip}
\vspace{\baselineskip}

\section*{Acknowledgements}
This work was partially supported by Wrocław University of Science and Technology statutory funds, the Polish National Science Centre, decisions no. 2013/09/B/ST6/02317, 2015/17/D/ST6/04046 and 2016/21/D/ST6/02408; the European Commission under the 7th Framework Programme, Coordination and Support Action, Grant Agreement Number 316097, ENGINE - European research centre of Network intelliGence for INnovation Enhancement (http://engine.pwr.edu.pl/); the RENOIR project - Reverse EngiNeering of sOcial Information pRocessing from the European Union’s Horizon 2020 research and innovation programme under the Marie Skłodowska-Curie grant agreement No. 691152, the Army Research Laboratory under Cooperative Agreement Number W911NF-09-2-0053 and by the Office of Naval Research Grant No. N00014-09-1-0607.

\section*{Author contributions}
JJ, PB created an initial concept of sequential seeding; JJ, PB, PK, BKS, RM developed the concept of sequential seeding to its current state; JJ, PB, PK, BKS designed the experiments; JJ, PB executed all experiments and simulations; BKS analyzed the proposed method coverage improvement theoretically; JJ, PB, PK, BKS, TK analyzed data and discussed results; JJ, PB, PK, RM drafted the manuscript. All authors critically reviewed the manuscript and approved the final version.

\end{document}